\documentclass[fleqn,twoside]{article}
\usepackage{espcrc2}
\usepackage{epsfig}
\usepackage{amsmath}
\usepackage{amssymb}
\usepackage{graphicx}
\newcommand{\bc}{\begin{center}}
\newcommand{\ec}{\end{center}}
\newcommand{\bq}{\begin{equation}}
\newcommand{\eq}{\end{equation}}
\newcommand{\bqa}{\begin{eqnarray}}
\newcommand{\eqa}{\end{eqnarray}}
\newcommand{\nn}{\nonumber}

\newcommand{\be}{\beta}

\newcommand{\ka}{\kappa}

\def\cO{{\cal O}}
\def\Source{{\Bbb S}}
\def\MSbar{{\rm  \overline{\footnotesize MS\kern-0.05em}\kern0.05em}}
\def\lrD{\stackrel{\leftrightarrow}{D}}

\title{
\vskip -110pt
{\small
\mbox{} \hfill DESY 04-190\\
\mbox{} \hfill SFB/CPP-04-54\\
\mbox{} \hfill September 2004\\}
\vskip 35pt
Recent results on moments of parton distribution functions
\thanks{Talk presented by I. Wetzorke. The work was supported 
by the DFG under SFB/TR 09-03.}}

\author{I.~Wetzorke\address[NIC]{NIC/DESY Zeuthen,
Platanenallee 6, D-15738 Zeuthen, Germany},
K.~Jansen\addressmark[NIC], 
F.~Palombi\address{E.~Fermi Research Center, c/o Compendio Viminale, pal.~F,
        I-00184 Rome, Italy},
A.~Shindler\addressmark[NIC]
}
       
\begin{document}

\begin{abstract}
We report on recent results for the pion and nucleon matrix element of
the twist-2 operator corresponding to the average momentum of
non-singlet quark densities. We discuss 
finite size effects for the nucleon matrix element and present first 
preliminary results for the non-perturbative renormalisation
from full dynamical simulations.
\vspace*{-2mm}\end{abstract}
\maketitle
\section{INTRODUCTION}\vspace*{-1mm}
Phenomenological fits to experimental data provide results for
moments of parton distribution functions which allow a direct comparison
with lattice calculations. In a recent paper \cite{ME04} we were able
to compute a continuum limit value of the lowest moment of a twist-2
operator, corresponding to the average momentum of non-singlet quark
densities, in pion states. 
In obtaining the renormalisation group invariant matrix element,
we have controlled important systematic errors, such as non-perturbative
renormalisation \cite{SSF03}, finite size effects \cite{FSE04} and effects of a
non-vanishing lattice spacing. The crucial limitation of our calculation was
the use of the quenched approximation.

In the scope of our current investigations is the comparison of finite size
effects for pion and nucleon matrix elements. In addition we concentrate on 
computing non-perturbatively the evolution of the twist-2 operators,
using the Schr\"odinger functional finite-size technique, on $N_f=2$ dynamical
configurations provided by the ALPHA collaboration \cite{ALPHA}.

\section{FINITE SIZE EFFECTS}\label{finite}\vspace*{-1mm}
Finite size effects for nucleon matrix elements have not been
studied in detail before, but might influence considerably the precise
determination of moments of parton distribution functions from lattice QCD.
We have investigated the effects of limited lattice extent at
$\be$=6.0 ($a\simeq0.093$ fm) with the non-perturbatively improved clover fermion
action and lattice sizes $(12-32)^3 \times 32$.

The twist-2 non-singlet operator for a flavour
doublet of fermions belongs to two irreducible
representations of the lattice $H(4)$ group
\bqa
{\cal O}_{12}({\rm x}) &=& \frac{1}{4}
       \bar\psi({\rm x}) \gamma_{\{1} \lrD_{2\}} \tau^3 \psi({\rm x})\nn\\
{\cal O}_{44} ({\rm x}) &=&  \frac{1}{2} \bar\psi({\rm x}) \Big[ \gamma_4 \lrD_4
- \frac{1}{3} \sum_{k=1}^3 \gamma_k \lrD_k \Big] \tau^3 \psi({\rm x})\;.\nn
\eqa
For the study of finite size effects (FSE) we concentrate solely on
the ${\cal O}_{44}$ operator,
since it can be computed at zero external momentum and thus provides
a reliable signal. In particular we use Schr\"odinger functional (SF)
boundary conditions \cite{Lues92} for our computation, which allows the
extraction of the matrix element from one large plateau around the middle
of the time extent $T$ of the lattice. 
The correlation function of the matrix element $f_M$ is obtained by inserting
the ${\cal O}_{44}$ operator between two SF nucleon states $\Source$ and
$\Source'$ at the time
boundaries $t=0$ and $T$ and suitable normalisation with the
boundary-to-boundary correlator $C_1$
\bqa
f_M(x_0/L) &=& - {1\over 2}\;\sum_{\rm x} %\sum_{\rm u,v,x,y,z} 
\langle \;\Source'%({\rm u},{\rm v}) 
\; {\cal O}_{44}({\rm x}) \;
\Source \; %({\rm y},{\rm z})
\rangle\nn\\
&\sim& e^{-m_N T} \; \langle N | {\cal O}_{44} | N\rangle
\; \{ 1 + \dots \}\nn\\[2mm]
C_1 &=& - {1\over 2}\;%\sum_{\rm u,v,y,z} 
\langle \;\Source'%({\rm u},{\rm v}) 
\Source \;%({\rm y},{\rm z}) 
\rangle \; \sim \; e^{-m_N T} \; .\nn
\eqa
The bare nucleon matrix element is defined by
\bqa
\langle x\rangle^{bare}_{u-d}
\equiv \frac{2\ka}{m_N} \langle N | {\cal O}_{44} | N\rangle
= \frac{2\ka}{m_N} f_M(x_0/L)/C_1 \; .\nn
\eqa

\begin{figure}[t]
\bc
\hskip-4mm
\epsfig{file=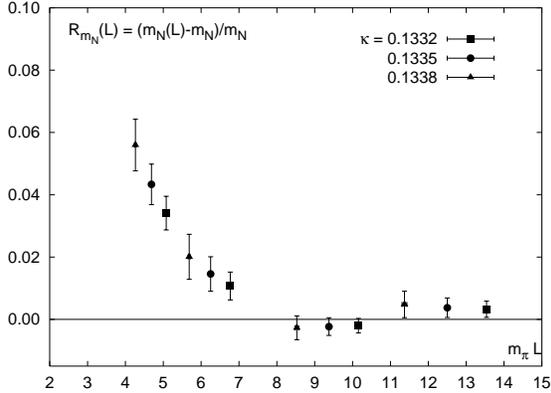,width=78mm}
\vskip-12mm
\ec
\caption{Finite size effects for $m_N a$\vspace*{-5mm}}
\label{f_N}
\end{figure}

\begin{figure}[t]
\bc
\hskip-4mm
\epsfig{file=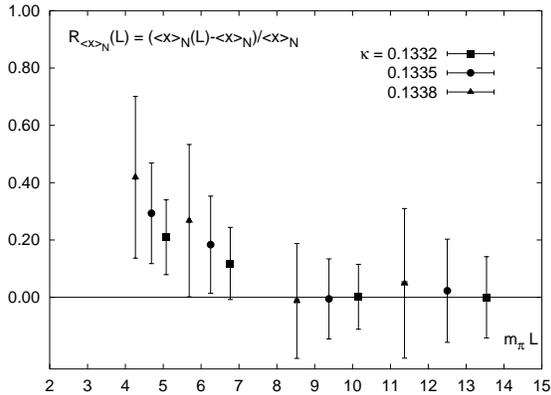,width=78mm}
\vskip-12mm
\ec
\caption{Finite size effects for nucleon $\langle x \rangle_{u-d}$\vspace*{-4.5mm}}
\label{f_X}
\end{figure}

In order to obtain the infinite volume limit nucleon mass and matrix element
we performed a purely phenomenological fit
$X(L)\!=\!c_0+c_1/L^{3/2}\exp(-c_2 L)$, where
$X\!\!=\!\!m_N a$ or $\langle x \rangle_{u-d}$.
Please note that a power-law fit
ansatz would describe the data almost equally well.
In figs.~\ref{f_N} and \ref{f_X} we show the relative systematic deviation
from the infinite volume limit values for both the nucleon mass and
matrix element. While the nucleon mass shows a deviation of about 4\% at
$m_\pi L \simeq 5$, the nucleon matrix element is overestimated by 10-30\%
at this value of $m_\pi L$. In both cases one can be sure to exclude FSE
only for $m_\pi L > 7$. For the pion matrix element we observed a FSE of
about 3\% at $m_\pi L \simeq 5$, while the pion mass is found to be free
from FSE already for $m_\pi L > 4$.

These observations lead to the expectation that the strong FSE for nucleon
matrix elements could be one ingredient for the deviation of current lattice
QCD estimates of $\langle x \rangle_{u-d}$ in comparison with
experimental data. 
\vspace*{-1mm}

\section{$N_f=2$ CONTINUUM STEP SCALING FUNCTION}\label{ssf}\vspace*{-1mm}
The strategy used to compute the non-perturbative evolution 
of the operators ${\cal O}_{44}$ and ${\cal O}_{12}$ has been
described in detail in our study for the quenched case \cite{SSF03}.
The evolution from initially large L/low $\mu$ to small L/high $\mu$
is obtained by applying the so called step scaling function (SSF) in the
Schr\"odinger functional scheme. Once the perturbative regime is safely
reached one continues the evolution in perturbation theory computing the 
scale and scheme independent RGI matrix element.
The connection with experimental results can be obtained by 
making the adequate perturbative evolution of the RGI matrix element
in the $\MSbar$ scheme.

\begin{figure}[t]
\bc
\hskip-4mm
\epsfig{file=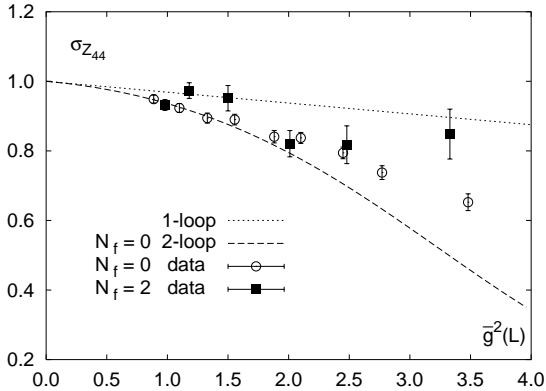,width=78mm}
\vskip-12mm
\ec
\caption{Continuum step scaling function $\sigma_{Z_{44}}$\vspace*{-3mm}}
\label{cont_X}
\end{figure}

The renormalisation conditions for the local ${\cal O}_{44}$ and 
${\cal O}_{12}$ operators are given by
\bq
{\cal O}_R(\mu) = Z_{\cal O}^{-1}(a\mu) {\cal O}(a), \quad {\cal O}_R(\mu = L^{-1}) = {\cal O}^{(0)} \;.\nn
\eq
The correlators to compute the Z factor are
\bq
f_{\cal O}(x_0/L) = - {1 \over 2}\;\sum_{\rm x,y,z}\;
\langle {\cal O}({\rm x})\;\Source_q({\rm y},{\rm z}) \rangle \nn
\label{eq:corr_func}
\eq
\bq
f_1 = - {1 \over 2}\;\sum_{\rm u,v,y,z}\;
\langle \Source'_q({\rm u},{\rm v}) \Source_q({\rm y},{\rm z})
\rangle \nn
\eq
where $\Source_q$ and $\Source_q'$ are suitable quark sources to probe the operators $\cO$.
With this definition the renormalisation constants are obtained by
\bq
Z(a/L,\mu) = c~ \frac{f_{\cal O}(x_0/L)}{\sqrt{f_1}} ; \quad 
c = \frac{\sqrt{f_1^{(0)}}}{f_{\cal O}^{(0)}(x_0/L)}. \nn
\eq
In order to map the $L$ dependence recursively we use the SSF, rigorously
defined on the lattice by
\bq
\sigma_{Z_{\cal O}} = \lim_{a \rightarrow 0} \Sigma_{Z_{\cal O}}(u,a/L)\nn
\eq
\bq
\Sigma_{Z_{\cal O}}(u,a/L) = \frac{Z_{\cal O}(u,2L/a)}{Z_{\cal O}(u,L/a)}, 
\quad u = \bar g^2_{SF}(L) \;.\nn
\eq
The values of $\beta$ corresponding to a fixed running coupling 
for $N_f=2$ dynamical $O(a)$-improved Wilson fermions are available in 
\cite{ALPHAc,ALPHA}. We have computed the SSF at six values of the 
renormalised coupling $\bar g^2_{SF}(L) = 0.98$ to $3.33$ for the lattice
sizes $L/a=6,8,12$, which includes the determination of Z factors on the
corresponding $L/a$ and $2L/a$ lattices. The continuum limit has been
obtained by a weighted average of several independent runs on the larger
$L/a=8,12$ (and $2L/a$) lattices using the HMC with one or two pseudo-fermion
fields \cite{GHMC}. In fig.~\ref{cont_X} we show the $N_f=2$ continuum
running of $\sigma_{Z_{44}}$ in comparison with the previously obtained
quenched evolution \cite{SSF03}. In addition we plot the universal 1-loop 
(dotted) and 2-loop order ($N_f = 0$, dashed) results of perturbation theory. 
The corresponding plot for the running of
$\sigma_{Z_{12}}$ shows a very similar behaviour.

Complemented by a computation of the pion or nucleon matrix element for
$N_f=2$ dynamical fermions, the SSF allows to obtain the RGI matrix element
completely non-perturbatively. The whole strategy is summarised by the
following formula:
\bqa
\langle \cO \rangle_{RGI} &=& \lim_{a\rightarrow 0}\frac{\langle \cO \rangle(a)}{Z_{\cal O}(a,\mu_0)} \times \nn \\
&& \sigma_{Z_{\cal O}}(\mu/\mu_0,\bar g^2(\mu)) {\cal F}_{SF}(\bar g^2(\mu)) \nn
\eqa
where we use the $n=6$ SSF computed at $\mu_n$
\bq
\sigma(\frac{\mu}{\mu_0},\bar g^2(\mu)) = 
\sigma(\frac{\mu_1}{\mu_0},\bar g^2(\mu_1)) \cdots \sigma(\frac{\mu_n}{\mu_{n-1}},\bar g^2(\mu_n)) \nn
\eq
to jump from the non-perturbative scale $\mu_0$ to the perturbative (ultraviolet)
scale $\mu$. At this point one can try to do the perturbative matching using
\bqa
{\cal F}_{SF}(\bar g^2(\mu)) &=& [\bar g^2(\mu)]^{-{\frac{\gamma_0}{2b_0}}} \times \nn \\ 
&& {\rm exp} \Big\{ -\int_0^{\bar g(\mu)} {\rm dx} \Big[ \frac{\gamma(x)}{\beta(x)} - \frac{\gamma_0}{2b_0}\Big]
\Big\} \nn
\eqa
computed with 3-loop $\beta$ and 2-loop $\gamma$ functions.
If the perturbative matching has been successful the quantity 
\bq
\mathfrak{\sigma}^{UV}_{INV}(\mu_0)=\sigma(\mu/\mu_0,\bar{g}^2(\mu))\;
{\cal F}_{SF}(\bar{g}^2(\mu))\nn
\eq
should be independent from the ultraviolet scale $\mu$. 
Indeed we found in the quenched case that the last steps give a very nice
plateau \cite{SSF03}. The determination of $\mathfrak{\sigma}^{UV}_{INV}$
for $N_f=2$ dynamical fermions requires a new computation of the 2-loop
anomalous dimension $\gamma_1^{SF}$ in the Schr\"odinger functional scheme,
which is currently work in progress.

\section*{ACKNOWLEDGEMENTS}\vspace*{-1mm}
The authors thank F.~Knechtli, M.~Della Morte and J.~Rolf
for the generation of dynamical configurations within the
running mass project \cite{ALPHA} and NIC/DESY Zeuthen for the
computer resources.

\end{document}